\documentclass[twoside]{article}
\usepackage{kyber2e}
\usepackage{graphicx}

\def \hs {\hspace*{-2mm}}
\newfont{\mi}{cmti9}
\newfont{\m}{cmr8}
\newfont{\ms}{cmsl8}
\newfont{\autor}{cmcsc10}
\newtheorem{theorem}{Theorem}

\newtheorem{lemma}{Lemma}

\begin{document}
% povinna titulni strana - Kybernetika
\rule[3mm]{128mm}{0mm}
\vspace*{-16mm}

{\footnotesize K\,Y\,B\,E\,R\,N\,E\,T\,I\,K\,A\, ---
\,V\,O\,L\,U\,M\,E\, {\it 4\,0\,} (\,2\,0\,0\,8\,)\,,\,
N\,U\,M\,B\,E\,R\, x\,,\,\  P\,A\,G\,E\,S\, \,x\,x\,x --
x\,x\,x}\\ \rule[3mm]{128mm}{0.2mm}

\vspace*{11mm}

{\large\bf \noindent 
ON\, THE\, SINGULAR\, LIMIT\, OF\, SOLUTIONS\, TO\, THE\, COX--INGERSOLL-ROSS\,  INTEREST\, RATE\, MODEL\, WITH\, STOCHASTIC\, VOLATILITY}

\vspace*{8mm}

{\autor \indent Be\'ata Stehl\'ikov\'a\, and\, Daniel \v{S}ev\v{c}ovi\v{c}
}\\

\vspace*{23mm}

\small
In this paper we are interested in term structure models for pricing zero coupon bonds under rapidly oscillating stochastic volatility. We analyze solutions to the generalized Cox--Ingersoll-Ross two factors model describing clustering of interest rate volatilities. The main goal is to derive an asymptotic expansion of the bond price with respect to a singular parameter representing the fast scale for the stochastic volatility process.  We derive the second order asymptotic expansion of a solution to the two factors generalized CIR model and we show that the first two terms in the expansion are independent of the variable representing stochastic volatility.
\begin{minipage}[t]{112mm}

\end{minipage} %\smallskip

\normalsize

\section{\hs INTRODUCTION}

Term structure models describe a functional dependence between the time to maturity of a discount bond and its present price. Yield of bonds, as a function of maturity, forms the so-called term structure of interest rates. If we denote by $P=P(t,T)$ the price of a bond paying no coupons at time $t$ with maturity at $T$ then the term structure of yields $R(t,T)$ is given by 
\[
P(t,T) = e^{-R(t,T)(T-t)},\quad \hbox{i.e.}\ R(t,T) = - \frac{\log P(t,T)}{T-t}
\]
(cf. Kwok \cite{kwok}). 
Continuous interest rate models are often formulated in terms of stochastic
differential equations (SDEs) for the instantaneous interest rate (or short rate)
as well as SDEs for other relevant quantities like e.g. volatility 
of the short rate process. In one-factor models there is a single 
stochastic differential equation for the short rate. The volatility of
the short rate process is given in a deterministic way. It is assumed to 
be constant (the Vasicek model \cite{vasicek}) or it is a function of the 
short rate itself. In the classical Cox, Ingersoll, and Ross model (CIR)  the short rate is modelled by a solution to the following stochastic differential equation:
\begin{equation}
dr = \kappa(\theta - r) dt + \sigma \sqrt{r} dw,
\label{CIR-short-rate}
\end{equation}
where $\kappa,\theta,\sigma>0$ are parameters representing the rate of reversion, the long term interest rate and volatility of the interest rate, respectively (see \cite{cir}). By $dw$ we have denoted the differential of the Wiener process. Beside these two 
simple models there is a wide range of other term structure models including, in
particular, the Chan-Karolyi-Longstaff-Sanders model \cite{ckls}, the Hull-White model \cite{h-w} 
and many others. Based on the assumption made on the form of the short rate process one can derive a linear scalar parabolic equation for the bond price as a function of the current short rate and time to maturity. 

In the two-factor models there are two sources of uncertainty yielding different term structures for the same short rate as they may depend on the value of the other factor. Moreover, two-factor models have a richer variety of possible shapes of term structures including, in particular, nonmonotone yield curves. The reader is referred to the books by Kwok \cite{kwok} and Brigo and Mercurio \cite{brigo-mercurio} for detailed discussion on two-factor interest rate modeling.

There are several ways how to incorporate the second stochastic factor.
It is reasonable to conjecture that in a financial market the volatility of a fluctuating underlying process for the short rate can be fluctuating as well. In the so-called two-factor models with a stochastic volatility we allow the volatility to have a stochastic behavior driven by another stochastic differential equation. 
As a consequence of the multidimensional It\=o's lemma
 the corresponding equation for the bond price
is a linear parabolic equation in two space dimensions. These spatial dimensions
correspond to the short rate and volatility. 
It is well known that the density distribution  of a stochastic process is a solution to the Focker-Planck partial differential equation
and can be expressed analytically in the case the volatility undergoes the Bessel square
root process (see e.g. \cite{goodman}). The actual value for the stochastic volatility is not know in the market. We can just observe its  statistical moments like e.g. the mean value, volatility, skewness of the volatility etc. 
Knowing the density distribution of the stochastic
volatility we are able to perform averaging of the bond price and the term structure
with respect to the stochastic volatility. Unlike the short rate which is known from the market data on daily 
basis, as it was already mentioned, the volatility of the short rate process is unknown. 
Therefore such a volatility averaging is of special importance for practitioners.

The main goal of the paper is to derive an asymptotic expansion of the bond price with respect to a singular parameter representing the fast scale for the stochastic volatility process. We derive the second order asymptotic expansion of a solution to the two factors generalized CIR model and we show that the first two terms in the expansion are independent of the stochastic volatility term. 

The paper is organized as follows. In the next section we present an empirical evidence of a short rate process for which the volatility is fluctuating and it has two concentration values. Next we discuss a model for statistical distribution capturing such a volatility clustering. Section 3 is devoted to the asymptotic analysis of solutions to the bond pricing equation in the case when the fluctuating volatility is rapidly oscillating. We derive explicit formulae for the first three terms in the asymptotic expansion of a solution with respect to a small parameter representing the fast time scale for rapidly oscillating volatility. 

%-----------------------------------------------------------------------
%                 SECTION: Volatility clustering
%-----------------------------------------------------------------------
\section{\hs Empirical evidence of existence of volatility clusters and their modeling}

\begin{figure}
\begin{center}
\includegraphics[height=3.8truecm]{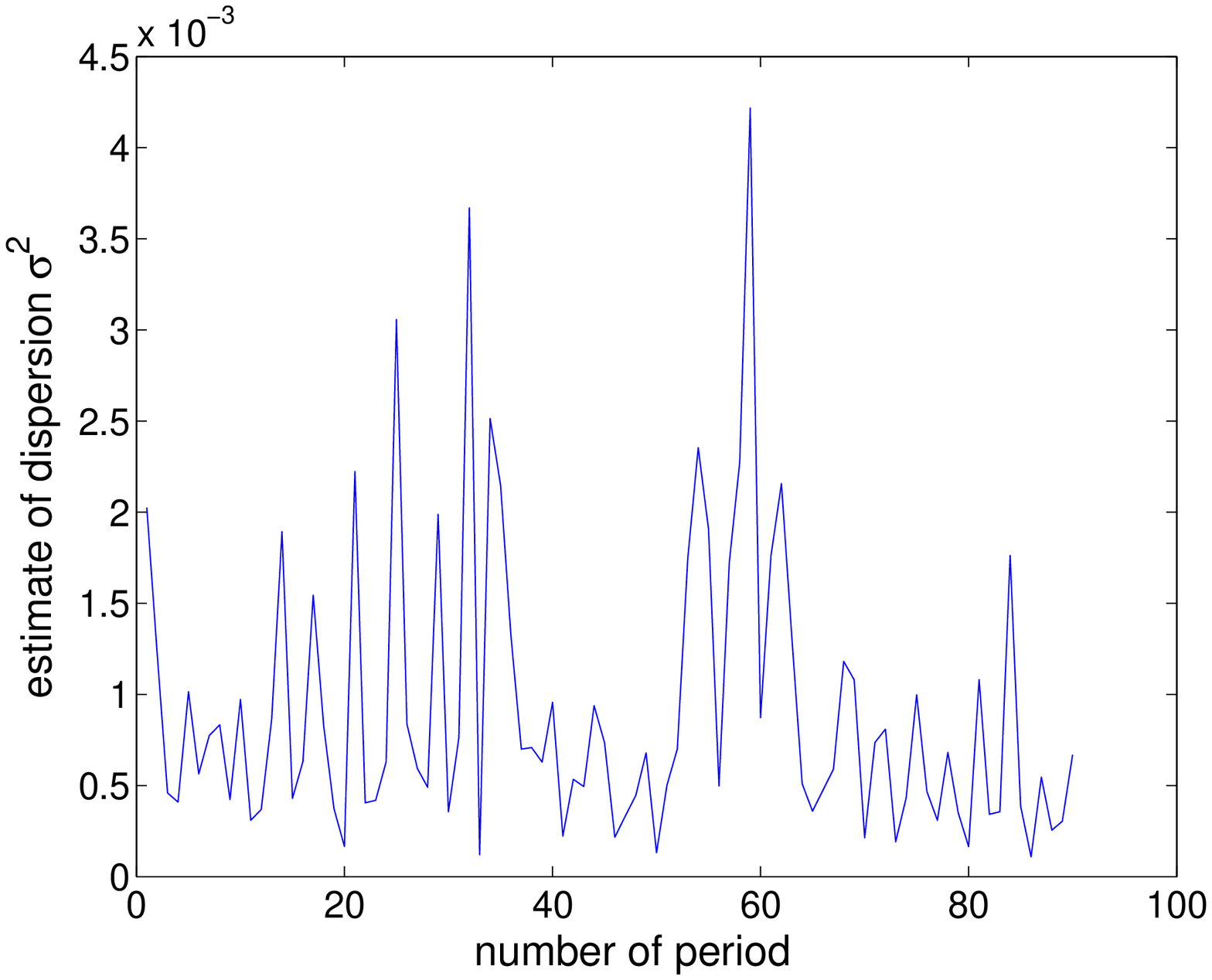}
\hglue1mm
\includegraphics[height=3.8truecm]{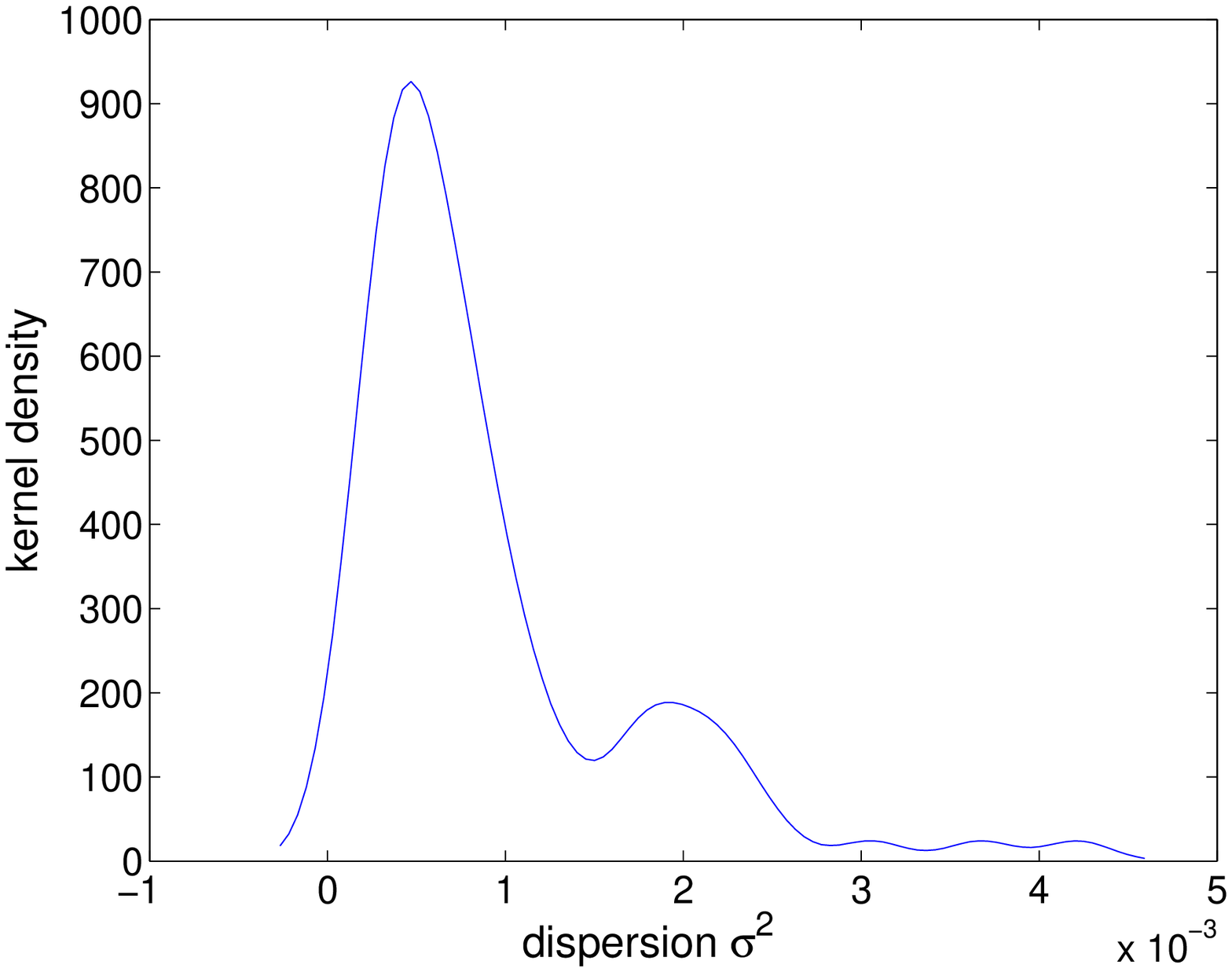}
\end{center}
\caption{\small Left: Estimates of CIR model's dispersion $\sigma^2$ from 20-day intervals (3-months Treasury bills, 90 intervals starting in January 1990). Right: the density distribution of estimates of the dispersion $\sigma^2$.}
\label{clustering}
\end{figure}

The key feature of the CIR modeling   consists in the assumption made on constant volatility of the stochastic process (\ref{CIR-short-rate}) driving the short rate $r$. However, in real financial markets we can observe a substantial deviation from this assumption. To provide an empirical evidence for such a volatility process, we computed maximum likelihood estimates of the dispersion for the CIR model for 20-day-long intervals using three months treasury bills data. Figure \ref{clustering} (left) shows the estimated dispersion as a function of time. Higher and lower volatility periods can be distinguished. They can be seen also on the kernel density estimates of the values in Figure \ref{clustering} (right).

\begin{figure} [ht]
\begin{center}
\includegraphics[width=0.4\textwidth]{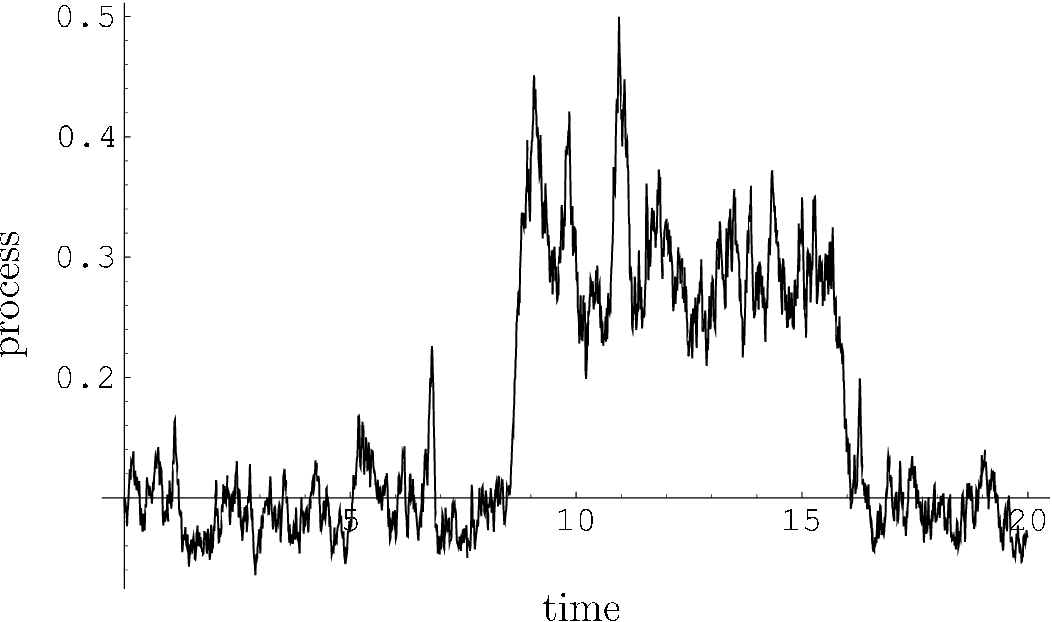}
\includegraphics[width=0.4\textwidth]{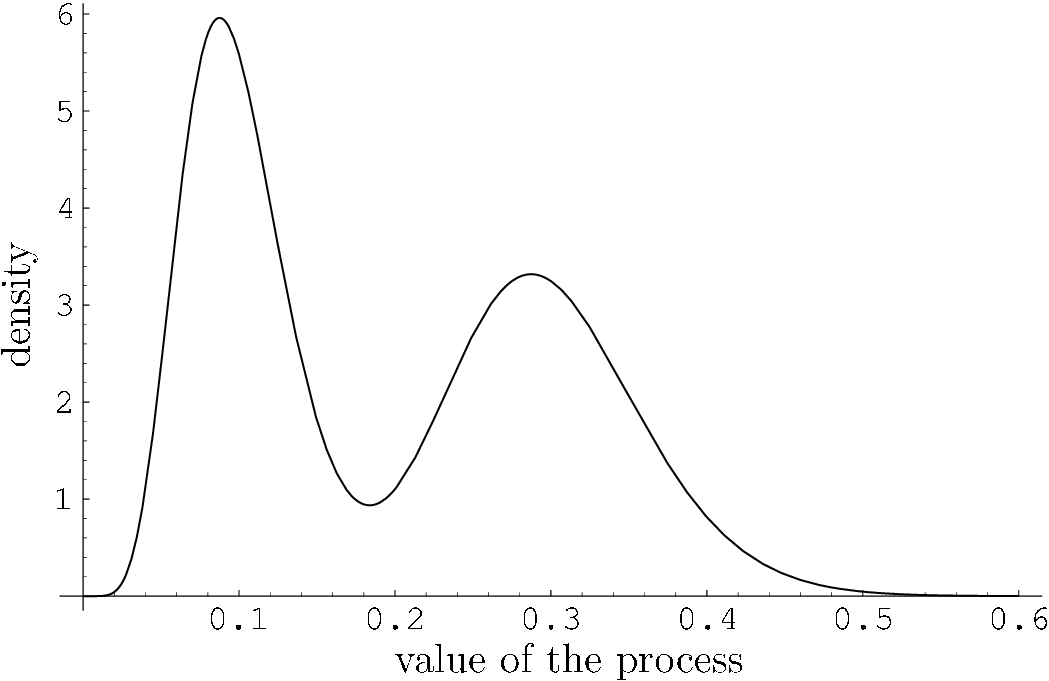}
\end{center}
\caption{\small Simulation of a process (left) and its asymptotic distribution (right).}
\label{obr-vol-clusters} 
\end{figure}

In order to capture such a behavior of the dispersion $\sigma^2$ we shall consider a model in which the limiting density of the dispersion (as $t\to\infty$) has two local maxima. It corresponds to the so-called volatility clustering phenomenon where the dispersion can be observed in the vicinity of two local maxima of the density distribution (see \cite{iscam05}). The desired behavior of the process and its limiting density are show in the Figure \ref{obr-vol-clusters}. A natural candidate for such a volatility process is
\begin{equation}
dy=\alpha(y)dt+\omega(y)dw
\label{general-proc}
\end{equation}
having a drift function $\alpha(y)$ such that the differential equation $\frac{dy}{dt}=\alpha(y)$ has two stable stationary solutions. With added stochastic part $\omega(y) dw$ of the process, these stationary solutions become values, around which the volatility concentrates. Recall that the cumulative distribution function $\tilde G=\tilde G(y,t)=Prob(y(t)<y|y(0)=y_0)$ of the process $y=y(t)$ satisfying (\ref{general-proc}) and starting almost surely from the initial datum $y_0$ can be obtained from a solution $\tilde g=\partial \tilde G/\partial y$ to the so-called Focker-Planck equation for the density function:
\begin{equation}
\frac{\partial \tilde g}{\partial t} = \frac12 \frac{\partial^2}{\partial y^2} (\omega(y)^2) \tilde g) - \frac{\partial}{\partial y}(\alpha(y) \tilde g) , \; \tilde g(y,0)=\delta(y-y_0)
\label{focker-planck}
\end{equation}
(cf. Kwok \cite{kwok}).
Here $\delta(y-y_0)$ denotes the Dirac delta function located at $y_0$. The limiting density $g(y)=\lim_{t\to\infty} \tilde g(y,t)$ of the process is therefore a stationary solution to the Focker-Planck equation (\ref{focker-planck}) and it forgets any information about the initial datum $y_0$, i.e 
\begin{equation}
L^*_0 g \equiv \frac12 \frac{\partial^2}{\partial y^2} (\omega(y)^2 g) - \frac{\partial}{\partial y}(\alpha(y)  g) =0\,.
\label{focker-planck-stat}
\end{equation}

In \cite{iscam05} one of the authors proposed a model with a property that the limiting density is a combination of two Gamma densities. Indeed, let us consider the following two mean reverting Bessel square root stochastic processes:
\begin{equation}
dy_i =  \kappa_y (\theta_i -y_i) dt+ v \sqrt{y_i} dw_i, \quad i=1,2,
\end{equation}
where $\theta_i >0$, $2 \kappa_y \theta_i > v^2>0$ for $i=1,2,$ and $dw_1, dw_2$ are uncorrelated differentials of the Wiener process. Solving the stationary Focker-Planck equation (\ref{focker-planck-stat}) it turns out that the limiting distributions of the processes $y_1,y_2$ are the Gamma distributions  with shape parameters $2 \kappa_y/v^2$ and $2 \kappa_y \theta_i/v^2$. Denote their densities by $g_1$ and $g_2$. Then $g_i(y)=C_i y^{\frac{2\kappa_y \theta_i}{v^2}-1} \exp(- \frac{2\kappa_y}{v^2}y)$ for $y>0$ and $g_i(y)=0$ otherwise. Here $C_i>0$ is a normalization constant such that $\int_{R} g_i(y) dy =1$.  Choose a parameter $k \in (0,1)$. Our aim is to construct a process with asymptotic density
\begin{equation}
g(y)=k g_1(y) + (1-k) g_2(y),
\label{gama-komb}
\end{equation}
corresponding to a convex mixture of densities $g_1$ and $g_2$. In the following theorem we see  that for the same square root volatility function of the form $v \sqrt{y}$ it is possible to achieve this goal. Drift of the process $\alpha(y)$ can be written as a weighted sum of drifts $\alpha_i(y)=\kappa(\theta_i-y), i=1,2,$ with the weights depending on $y$. 

\begin{theorem}
\label{clustering-vahy}
\cite[Section 5]{iscam05}
Suppose that the drift term $\alpha$ has the form: $\alpha(y)=w(y) \alpha_1(y) + (1-w(y)) \alpha_2(y)$ where 
$w(y)=k g_1(y)/(k g_1(y) + (1-k) g_2(y))$ and $\alpha_i(y)=\kappa(\theta_i-y)$. Then the stochastic process driven by the SDE: 
$dy=\alpha(y)dt+ v \sqrt{y} dw$ has the limiting distribution $g$ given by the convex combination (\ref{gama-komb}) of densities $g_1, g_2$.
\end{theorem}

%-----------------------------------------------------------------------
%                 SECTION: Fast time scale for volatility
%-----------------------------------------------------------------------

\section{\hs Generalized CIR model with rapidly oscillating stochastic volatility and its asymptotic analysis}

The aim of this section is to provide a tool for modeling the effects of rapidly oscillating stochastic volatility that can be observed in real markets. If the length of the time scale for dispersion $y$ is denoted by $\varepsilon$, the equation for the variable $y$ reads as follows:
\begin{equation}
dy=\frac{\alpha(y)}{\varepsilon}dt + \frac{v \sqrt{y}}{\sqrt{\varepsilon}} dw_y.
\label{slowly-osc}
\end{equation}
In what follows we will assume that $0< \varepsilon \ll 1$ is a small singular parameter.
Notice that the limiting density function $g$ of the stochastic process driven by SDE (\ref{slowly-osc}) is independent of the scaling parameter $\varepsilon>0$. The statement follows directly from the stationary Focker-Planck equation (\ref{focker-planck-stat}). 
Concerning structural assumption made on the drift function $\alpha:R\to R$ we shall henceforth assume the following hypothesis:
\[
(A)\qquad\qquad 
\alpha \ \ \hbox{is a } C^1 \ \hbox{function on } [0,\infty),\ \  \frac{2\alpha(0)}{v^2} >1,\ \ \limsup_{y\to\infty} \frac{\alpha(y)}{y} <0.
\]

Now it is straightforward computation to verify the following auxiliary lemma. 

\begin{lemma}
\label{mixture}
Let the drift function $\alpha(y)$ be defined as a mixture of two Gamma limiting distributions as in Theorem~\ref{clustering-vahy}. Then the function $\alpha$ satisfies the hypothesis (A) with $\alpha(0)=\kappa\min(\theta_1,\theta_2)$ and $\limsup_{y\to\infty} \frac{\alpha(y)}{y}=-\kappa<0$.
\end{lemma}

Next we shall show the limiting density $g$ of the process driven by SDE (\ref{slowly-osc}) is uniquely given by the following lemma:

\begin{lemma}
\label{density}
Under the hypothesis (A) made on the drift function $\alpha$ the stationary Focker-Planck equation $L^*_0 g \equiv \frac{v^2}{2}\frac{\partial^2}{\partial y^2} (y g) - \frac{\partial}{\partial y}(\alpha(y)  g) =0$ has a unique solution $g$ such that $g(0) = 0$ for $y \leq 0$. It can be explicitly expressed as:
\[
g(y) = C y^{-1}\exp\left(\frac{2}{v^2} \int_1^y \frac{\alpha(\xi)}{\xi} d\xi\right) = C y^{\frac{2\alpha(0)}{v^2} -1} \exp\left(\frac{2}{v^2} \int_1^y \hat\alpha(\xi) d\xi\right)
\]
for $y>0$ and $g(y)=0$ for $y \leq 0$. Here $\hat\alpha(y)= (\alpha(y)-\alpha(0))/y$ and $C>0$ is a normalization constant such that $\int_0^\infty g(y) dy =1$. 
\end{lemma}
{\sl Proof.} It follows by direct verification of the equation. The other linearly independent solution $g_2$ to the  equation (\ref{focker-planck-stat})  has a nontrivial limit $g_2(0+)>0$. \hfill $\diamondsuit$

In what follows, we denote by $\sigma^2, D>0,$ and $S$ the limiting mean value, dispersion and skewness of the stochastic process for the $y$-variable representing stochastic dispersion, i.e. 
\begin{equation}
\sigma^2 = \int_0^\infty y g(y)\, dy, \quad D  = \int_0^\infty (y-\sigma^2)^2 g(y)\, dy,\quad 
S= \frac{1}{D^{\frac32}}\int_0^\infty (y-\sigma^2)^3 g(y) dy\,.
\label{sigma}
\end{equation}
Notice that $D = - \int_0^\infty \int_0^y (\xi-\sigma^2) g(\xi)\, d\xi dy$.
In the generalized CIR model with a stochastic volatility, the instantaneous interest rate (short rate) $r$ will be modelled by the mean reverting process of the form (\ref{CIR-short-rate}) where the volatility of is replaced by a square root of a stochastic dispersion $y$, i.e. 
\begin{equation}
dr = \kappa(\theta - r) dt + \sqrt{y} \sqrt{r} dw_r\,.
\label{CIR-short-rate-stoch}
\end{equation}
The differentials of the Wiener processes $dw_y$ and $dw_r$ are assumed to be uncorrelated throughout the paper, i.e. $E(dw_y dw_r) = 0$. Then the corresponding partial differential equation for the bond price $P^\varepsilon= P^\varepsilon(t,r,y)$ has the following form:
\begin{eqnarray}
\frac{\partial P^\varepsilon}{\partial t} &+& (\kappa (\theta-r) - \tilde\lambda_1(y,r) r^{\frac12} \sqrt{y} ) \frac{\partial P^\varepsilon}{\partial r} + \frac{1}{2} r y \frac{\partial^2 P^\varepsilon}{\partial r^2} - r P^\varepsilon \\
&+& \frac{1}{\sqrt{\varepsilon}} \left(  - \tilde\lambda_2(y,r) v \sqrt{y} \frac{\partial P^\varepsilon}{\partial y}\right) + \frac{1}{\varepsilon} \left(\alpha(y) \frac{\partial P^\varepsilon}{\partial y} + \frac {v^2 y}{2} \frac{\partial^2 P^\varepsilon}{\partial y^2}\right) = 0, \nonumber
\label{pde}
\end{eqnarray}
$(t,r,y)\in Q_T\equiv(0,T)\times R^+\times R^+,$
where $\tilde{\lambda}_1,\tilde{\lambda}_2$ are the so-called market prices of risk (cf. \cite[Chapter 7]{kwok}).
By a solution $P^\varepsilon$ to (\ref{pde}) we mean a bounded function $P^\varepsilon\in C^{1,2}(Q_T)\cap C(\bar Q_T)$ satisfying equation (\ref{pde}) on $\bar Q_T$.   Concerning the structural form of market prices of risk functions $\tilde{\lambda}_1,\tilde{\lambda}_2$ we shall suppose that 
\[
\tilde\lambda_1(t,r,y)=\lambda_1 \sqrt{r}\sqrt{y}, \quad \tilde\lambda_2(t,r,y)=\lambda_2 \sqrt{y}
\]
where $\lambda_1,\lambda_2\in R$ are constants. It is worthwile noting that the latter assumption is not restrictive as the original one-factor CIR model assumes such a form of the market price of risk (cf. Kwok \cite{kwok}). We shall rewrite PDE (\ref{pde}) in the operator form:
\begin{equation}
(\varepsilon^{-1} \mathcal{L}_0 + \varepsilon^{-1/2} \mathcal{L}_1 + \mathcal{L}_2 ) P^{\varepsilon} = 0,
\label{eq:bondprice2}
\end{equation}
where the linear differential operators $\mathcal{L}_0, \mathcal{L}_1,  \mathcal{L}_2$ are defined as follows:
\begin{eqnarray}
\mathcal{L}_0&=&\alpha(y) \frac{\partial }{\partial y} + \frac {v^2 y}{2} \frac{\partial^2 }{\partial y^2},  \;
\mathcal{L}_1=   - \lambda_2 v y \frac{\partial }{\partial y},  \nonumber \\
\mathcal{L}_2&=&\frac{\partial }{\partial t} + (\kappa (\theta-r) - \lambda_1 r y ) \frac{\partial }{\partial r} + \frac{1}{2} r y \frac{\partial^2 }{\partial r^2} - r. 
\end{eqnarray}
Next we expand the solution $P^{\varepsilon}$ into Taylor power series:
\begin{equation}
\label{expansion}
P^{\varepsilon}(t,r,y)= \sum_{j=0}^\infty \varepsilon^{\frac{j}{2}}  P_j(t,r,y)
\end{equation}
with the terminal conditions $P_0(T,r,y)=1, P_j(T,r,y)=0 \; \textrm{for } j \geq 1$ at expiry $t=T$.
The main goal of this paper is to examine the singular limiting behavior of a solution $P^\varepsilon$ as $\varepsilon\to 0^+$. More precisely, we shall determine the first three terms $P_0,P_1,P_2$ of the asymptotic expansion (\ref{expansion}). 
We shall henceforth denote by $\langle\psi\rangle$ the averaged value of the function $\psi\in C([0,\infty))$ with respect to the density $g$, i.e. $\langle\psi\rangle = \int_0^\infty \psi(y)g(y)\,dy$. We shall also use the notation $\langle \mathcal{L}_2\rangle$ standing for the averaged linear operator $\mathcal{L}_2$, i.e. 
\begin{equation}
\langle \mathcal{L}_2\rangle
\equiv 
\frac{\partial }{\partial t}+ \left( \kappa(\theta-r) - \lambda_1 r \sigma^2  \right) \frac{\partial }{\partial r}+\frac{1}{2} \sigma^2 r  \frac{\partial^2 }{\partial r^2} - r 
\label{averaged-L2}
\end{equation}

\begin{lemma}
\label{L0-average}
Let $\psi\in C^1([0,\infty))$ be such that $\mathcal{L}_0 \psi$ is bounded. Then  $\langle\mathcal{L}_0\psi\rangle = 0$.
\end{lemma}
{\sl Proof.} Notice that the operator $\mathcal{L}_0^*$ is the adjoint operator to the linear operator $\mathcal{L}_0$ with respect to the $L^2$--inner product $(\psi,\phi) = \int_0^\infty \psi(y)\phi(y)\,dy$. It means that $\langle\mathcal{L}_0\psi\rangle = (\mathcal{L}_0\psi, g) = 
(\psi,\mathcal{L}_0^* g)=0$ because the density $g$ is a solution to Eq. (\ref{focker-planck-stat}). \hfill $\diamondsuit$

\smallskip
The following lemma will be useful when computing higher order term in series expansion (\ref{expansion}).

\begin{lemma}
\label{L0-equation}
Let $F \in C([0,\infty))$ be such that $\langle F\rangle =0$. Then, up to an additive constant, there exists a unique solution $\psi\in C^2((0,\infty))\cap C([0,\infty))$ to the non-homogeneous equation
$\mathcal{L}_0 \psi = \frac{v^2}{2}F$.  Its derivative $\frac{\partial\psi}{\partial y}$ is given by
\[
\frac{\partial \psi}{\partial y}(y) = \frac{1}{yg(y)} \int_0^{y} F(\xi) g(\xi) d\xi.
\]
Moreover, $\langle\mathcal{L}_1 \psi \rangle = \lambda_2 v \int_0^\infty F(y) y g(y) dy$. 
In particular, if $\psi$ is a solution to the equation $\mathcal{L}_0 \psi =0$ then $\psi$ is a constant function with respect to the $y$-variable.
\end{lemma}
{\sl Proof.} Using equation (\ref{focker-planck-stat}) for the limiting density $g$ and inserting $\frac{\partial \psi}{\partial y}$ into the operator $\mathcal{L}_0$ we obtain that $\psi$ is a solution to the equation $\mathcal{L}_0 \psi = \frac{v^2}{2}F$. Other independent solutions are not continuous at $y=0$. The formula for $\langle\mathcal{L}_1 \psi \rangle$ follows from the definition of the operator $\mathcal{L}_1$ by applying integration by parts formula.\hfill $\diamondsuit$

\medskip
\noindent Now we proceed with collecting the terms of the power series expansion of (\ref{eq:bondprice2}). 

\smallskip
\noindent $\bullet$ In the order $\varepsilon^{-1}$ we have $\mathcal{L}_0 P_0=0$. According to Lemma~\ref{L0-equation} we have $P_0=P_0(t,r)$, i.e. $P_0$ is independent of the $y$-variable.

\smallskip
\noindent $\bullet$  In the order $\varepsilon^{-1/2}$ we have $\mathcal{L}_0 P_1 + \mathcal{L}_1 P_0 =0$. Since $P_0=P_0(t,r)$ we deduce $\mathcal{L}_1 P_0=0$ and so $\mathcal{L}_0 P_1=0$. By Lemma~\ref{L0-equation}, $P_1=P_1(t,r)$ is independent of $y$.

\smallskip
\noindent $\bullet$  In the order $\varepsilon^{0}$ we have $\mathcal{L}_0 P_2 + \mathcal{L}_1 P_1 + \mathcal{L}_2 P_0 =0$. Since $P_1=P_1(t,r)$ we have $\mathcal{L}_1 P_0=0$. Hence 
$\mathcal{L}_0 P_2 + \mathcal{L}_2 P_0 =0$. Taking the average $\langle.\rangle$ of both sides of the latter equation we obtain $\langle \mathcal{L}_0 P_2 \rangle + \langle \mathcal{L}_2 P_0 \rangle =0$. By Lemma~\ref{L0-average} and the fact that $P_0$ is independent of $y$-variable we conclude $\langle \mathcal{L}_2 \rangle  P_0 = \langle \mathcal{L}_2 P_0 \rangle  = 0$. Therefore $P_0$ is a solution to the classical one-factor PDE equation for the CIR model satisfying the terminal condition $P_0(T,r)=1$ for any $r\ge 0$. 
It is well known that the solution $P_0=P_0(t,r)$ to the  equation $\langle \mathcal{L}_2 \rangle P_0 = 0$ is given by the explicit formula:
\begin{equation}
P_0(t,r) = A_0(t) e^{-B(t) r},
\label{CIR-explicit}
\end{equation}
where $A_0^\prime = \kappa\theta B$ and $B^\prime = (\kappa+\lambda_1\sigma^2) B +\frac{\sigma^2}{2}B^2 -1$,
$A_0(T)=1, B(T)=0,$ i.e. 
\[
A_0(t) = \left( \frac{2 \phi e^{(\phi+\psi)(T-t)/2}} {(\phi+\psi)(e^{\phi(T-t)}-1)+2\phi} \right)^{\frac{2\kappa \theta}{\sigma^2}},
\  B(t) =\frac{2(e^{\phi(T-t)}-1)}{(\phi+\psi)(e^{\phi(T-t)}-1)+2\phi},
\]
$\psi=\kappa+\lambda_1 \sigma^2, \; \phi=\sqrt{\psi^2+2\sigma^2}$ (cf. Kwok \cite[Chapter 7]{kwok}). Since $\langle \mathcal{L}_2\rangle P_0 = 0$  we have
$ -\mathcal{L}_2 P_0 = \left( \langle \mathcal{L}_2\rangle - \mathcal{L}_2 \right)P_0 = (\sigma^2-y) f(t) r e^{-B(t) r}$ where
\[
f(t)= (\lambda_1 B(t) + \frac{1}{2}B(t)^2 ) A_0(t).
\]
Hence $\mathcal{L}_0 P_2 = - \mathcal{L}_2 P_0 = (\sigma^2-y) f(t) r e^{-B(t) r}$. According to Lemma~\ref{L0-equation} we have 
\begin{equation}
\frac{\partial P_2}{\partial y} = -\frac{2}{v^2} f(t) r e^{-B(t) r}  H(y), \ \ 
H(y) = \frac{1}{yg(y)} \int_0^{y} (\xi-\sigma^2) g(\xi) d\xi.
\label{derP2}
\end{equation}

\begin{figure}
\begin{center}
\includegraphics[width=0.48\textwidth]{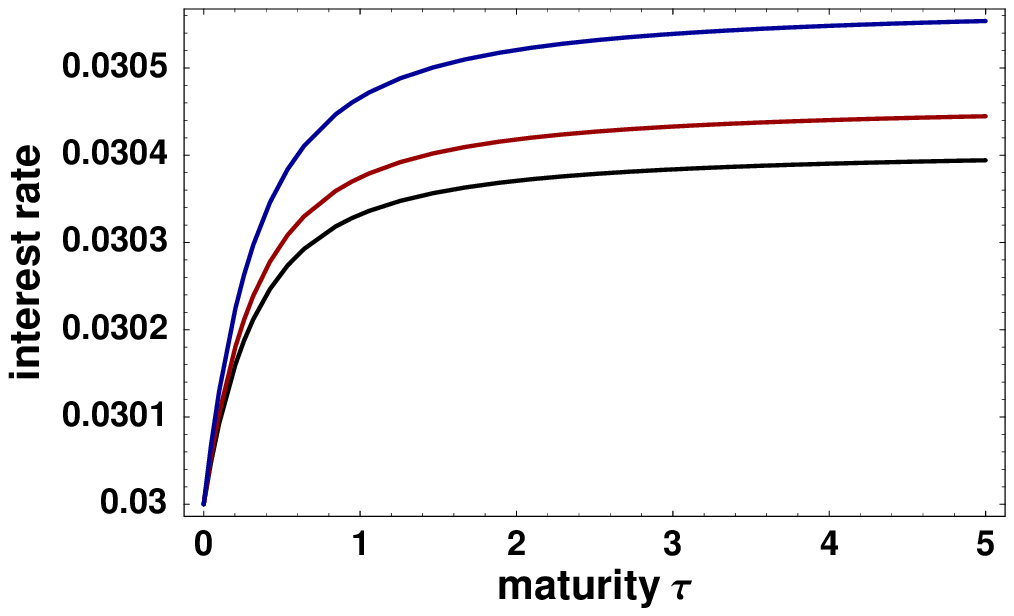}
\includegraphics[width=0.48\textwidth]{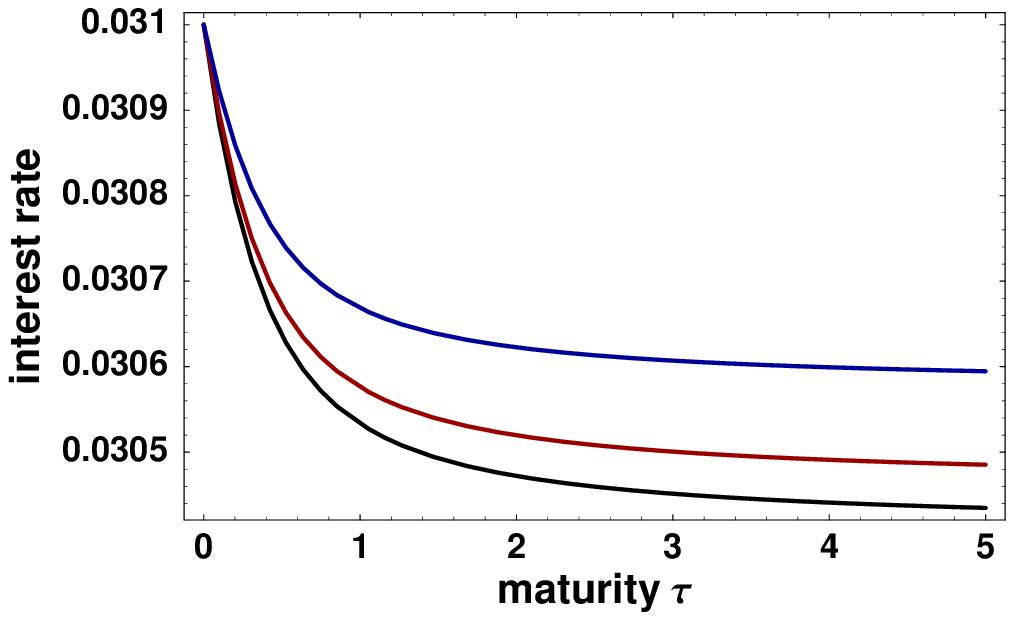}
\end{center}
\caption{\small The approximate term structure 
$R^\varepsilon=R^\varepsilon(T-\tau,r)$ based on the first two leading terms of 
the bond price $P^\varepsilon \approx P_0(T-\tau,r) +\sqrt{\varepsilon}
P_1(T-\tau,r)$ starting from the short rate $r=0.03$ (left) and
$r=0.031$ (right) for several values of the singular parameter $\varepsilon = 0, 0.001,0.01$  (black, red and blue curves), resp.}
\label{obr-epsilon} 
\end{figure}

\smallskip
\noindent $\bullet$ In the order $\varepsilon^{1/2}$ we have  $\mathcal{L}_0 P_3 + \mathcal{L}_1 P_2 + \mathcal{L}_2 P_1 =0$. Since $\langle \mathcal{L}_0 P_3 \rangle =0$ we have $\langle \mathcal{L}_1 P_2 \rangle + \langle \mathcal{L}_2 P_1 \rangle =0$. The function $P_1=P_1(t,r)$ is independent of the $y$-variable and therefore $\langle \mathcal{L}_2 \rangle P_1 = \langle \mathcal{L}_2 P_1 \rangle = - \langle \mathcal{L}_1 P_2 \rangle$.
By Lemma~\ref{L0-equation} we have 
\[
\mathcal{L}_1 P_2= \frac{2\lambda_2}{v} f(t) r e^{-B(t) r}  yH(y),\quad -\langle \mathcal{L}_1 P_2 \rangle = K_1 f(t) r e^{-B(t)r},
\]
where $K_1= -\frac{2 \lambda_2}{v} \int_0^{\infty} \int_0^y (\xi-\sigma^2) g(\xi) d\xi dy = \frac{2 \lambda_2}{v} D$ is a constant (see (\ref{sigma})). Notice that the constant $K_1$ and the function $f(t)$ depend on the first two moments $\sigma^2$ and $D$ of the stochastic dispersion only. Equation $\langle \mathcal{L}_2 \rangle P_1 = \langle \mathcal{L}_2 P_1 \rangle = - \langle \mathcal{L}_1 P_2 \rangle$ reads as:
\begin{equation}
\frac{\partial P_1}{\partial t}+ \left( \kappa(\theta-r) -\lambda_1 r \sigma^2 \right) \frac{\partial P_1}{\partial r}+\frac{1}{2} r \sigma^2  \frac{\partial^2 P_1}{\partial r^2} - r P_1 = K_1 f(t) r e^{-B(t)r}.
\label{P1-equation}
\end{equation}
The solution $P_1$ satisfying the terminal condition $P_1(T,r)=0$ for $r \geq 0$ can be found in the closed form:
\begin{equation}
P_1(t,r)=(A_{10}(t)+A_{11}(t)r)e^{-B(t)r}
\label{P1-formula}
\end{equation}
where the functions $A_{10}(t)$, $A_{11}(t)$ are solutions to the system of linear ODEs:
\begin{eqnarray}
A^\prime_{11}(t)&=& \left(\kappa \theta B(t) + \kappa + \lambda_1 \sigma^2 + \sigma^2 B(t)\right) A_{11}(t) + K_1 f(t), \\
A^\prime_{10}(t)&=& \kappa \theta B(t) A_{10}(t) - \kappa \theta A_{11}(t), \nonumber
\label{ODE}
\end{eqnarray}
with terminal conditions $A_{10}(T)=0$, $A_{11}(T)=0$. We can analytically and also numerically compute $A_{10}$, $A_{11}$ in a fast and accurate manner. This way we have obtained the term $P_1(t,r)$. In Figure~\ref{obr-epsilon} examples of numerical approximation of the term structure 
$R^\varepsilon(T-\tau,r) = - \frac{1}{\tau}\log \langle P^\varepsilon(T-\tau,r,.)\rangle$ corresponding to the second order expansion of the averaged value of $\langle P^\varepsilon(t,r,.)\rangle$,  $P^\varepsilon(t,r,y) \approx P_0(t,r) +\sqrt{\varepsilon} P_1(t,r)$. We plot term structures starting from the short rate $r=0.03$ (left) and
$r=0.031$ (right) for parameters $\kappa=5$, $\theta=0.03$, $\kappa_y=100$, $v=1.1832$, $\theta_1=0.025$, $\theta_2=0.1$, $k=1/3$, $\lambda_1=-1$, $\lambda_2=-100$ and $\varepsilon = 0, 0.001,0.01$  (black, red and blue curves).

\smallskip
Having $P_1$ and $\frac{\partial P_2}{\partial y}$ we can compute the term 
$\mathcal{L}_1 P_2 + \mathcal{L}_2 P_1$. With regard to Lemma~\ref{L0-equation} equation $\mathcal{L}_0 P_3 = -\mathcal{L}_1 P_2 - \mathcal{L}_2 P_1$ then yields a  formula for  $\frac{\partial P_3}{\partial y}$ and 
\[
\langle \mathcal{L}_1 P_3 \rangle = -\frac{2\lambda_2}{v}\left( ( \frac{2\lambda_2}{v} K_3 +K_1\sigma^2) f(t)r e^{-B(t)r} + D ( -\lambda_1 r\frac{\partial P_1}{\partial r} + \frac{r}{2}\frac{\partial^2 P_1}{\partial r^2})\right)
\]
where the constant $K_3=\int_0^\infty \xi^3 H(\xi) g(\xi) d\xi = - \frac{1}{2} S D^{\frac32} - \sigma^2 D$ depends on the first three statistical moments of the stochastic dispersion. 

\noindent $\bullet$ In the order $\varepsilon^{1}$ we have  $\mathcal{L}_0 P_4 + \mathcal{L}_1 P_3 + \mathcal{L}_2 P_2  = 0$. Proceeding similarly as before we have  $\langle \mathcal{L}_0 P_4 \rangle =0$ and therefore 
\begin{equation}
\langle \mathcal{L}_1 P_3 \rangle + \langle \mathcal{L}_2 P_2 \rangle = 0.
\label{eq}
\end{equation}
We decompose  the function $P_2(t,r,y)$ in the form
\begin{equation}
P_2(t,r,y)=\bar{P}_2(t,r)+\tilde{P}_2(t,r,y), \label{p2}
\end{equation}
where $\bar{P}_2$ is the averaged value of $P_2$ and $\tilde{P}_2$ is a zero mean fluctuation, i.e. $\langle \tilde{P}_2 \rangle =0$. As $\bar{P}_2$ does not depend on $y$, we have $\frac{\partial \tilde{P}_2}{\partial y} = \frac{\partial P_2}{\partial y}$. Taking into account  $\langle \mathcal{L}_2 \tilde P_2 \rangle = 0$ 
we obtain
\[
\tilde{P}_2(t,r,y)=  -\frac{2}{v^2} f(t) r e^{-B(t)r} \left(
\int_0^y H(\xi) d\xi - K_2
\right)
\]
where $K_2= \int_0^{\infty} g(s) \int_0^s H(\xi) d\xi ds$ is a constant and  the function $H$ is given by (\ref{derP2}).
Now we can use decomposition (\ref{p2}) to evaluate $\langle \mathcal{L}_2 P_2 \rangle$. We have 
$\langle \mathcal{L}_2 P_2 \rangle = \langle \mathcal{L}_2 (\bar{P}_2 + \tilde{P}_2 ) \rangle =
\langle  \mathcal{L}_2  \rangle \bar{P}_2 + \langle  \mathcal{L}_2 \tilde{P}_2 \rangle$ because $\bar{P}_2$ is independent of $y$.
Next we can determine $\langle \mathcal{L}_2 \tilde P_2 \rangle$ in the following form:
\[
\langle \mathcal{L}_2 \tilde P_2 \rangle = -\frac{2}{v^2} K_4 f(t) r \left( -\lambda_1 \frac{\partial}{\partial r} + \frac12 \frac{\partial^2}{\partial r^2} \right) (r e^{-B(t) r})
\]
where $K_4=\int_0^\infty \int_0^y H(\xi)d\xi (y-\sigma^2) g(y) dy$. It is worthwile noting that both constants $K_2,K_4$ depend on all nontrivial statistical moments of the stochastic dispersion. Equation (\ref{eq}) then becomes
\[
\langle  \mathcal{L}_2 \rangle \bar{P}_2 = -\langle  \mathcal{L}_2 \tilde{P}_2 \rangle - \langle  \mathcal{L}_1 P_3 \rangle = (a(t) + b(t) r + c(t) r^2) e^{-B(t) r}, \; \bar{P}_2(T,r)=0,
\]
which is a partial differential equation for $\bar{P}_2=\bar{P}_2(t,r,y)$ with a right hand side which can be explicitly computed from already obtained results in the closed form:
\begin{equation}
\bar P_2(t,r) = (A_{20}(t) + A_{21}(t) r + A_{22}(t) r^2) e^{-B(t) r}
\label{P2-equation}
\end{equation}
where the functions $A_{20}, A_{21}, A_{22}$ are solutions to a linear system of ODEs. We omit details here.

In summary we have shown the following main result of this paper:

\begin{theorem}
\label{veta}
The solution $P^\varepsilon = P^\varepsilon(t,r,y)$ of the generalized CIR bond pricing equation (\ref{pde}) with rapidly oscillating dispersion can be approximated, for small values of the singular parameter $0<\varepsilon\ll 1$, by $P^\varepsilon(t,r,y) \approx P_0(t,r) +\sqrt{\varepsilon} P_1(t,r) + \varepsilon P_2(t,r,y) + O(\varepsilon^\frac32)$. 

The first two terms $P_0,P_1$ are independent of the $y$-variable representing unobserved stochastic volatility. They depend only the first two statistical moments (mean value and dispersion) of the stochastic dispersion and other model parameters. 

The next term in the expansion $P_2$ nontrivially depends on the $y$-variable. $P_2$ as well as its averaged value $\langle P_2\rangle$ depends also on all nontrivial statistical moments of the stochastic dispersion. 

The terms $P_0,P_1,P_2$ can be evaluated by closed-form formulae (\ref{CIR-explicit}), (\ref{P1-formula}), (\ref{P2-equation}). 
\end{theorem}

\normalsize
\section*{ACKNOWLEDGMENT}
\small
The support from the grant VEGA 1/3767/06 and is kindly acknowledged.

\footnotesize
\begin{flushright}
(Received \today)\,\ \rule{0mm}{0mm}
\end{flushright}

\vspace*{2mm}

{\mi
\begin{flushright}
\begin{minipage}[]{124mm}
{
Be\'ata Stehl\'\i kov\'a and  Daniel  \v Sev\v covi\v c, 
Department of Applied Mathematics and Statistics, Comenius University, Mlynsk\' a dolina, 842 48 Bratislava, Slovakia
\\ e-mails: 
\{stehlikova,sevcovic\}@fmph.uniba.sk
}
\end{minipage}
\end{flushright}
}

\end{document}